\documentclass[12pt]{iopart}
\usepackage {setstack}
\usepackage{iopams}
\usepackage {amscd}
\usepackage {latexsym}
\usepackage {amssymb}
\usepackage {amsfonts}
\usepackage {graphics}
\usepackage {graphicx}

\newcommand {\oks}[2]{{\raise0.7ex\hbox{${\scriptstyle #1}$}\!\mathord{\left/
{\vphantom{{1}{2}}}\right.\kern-\nulldelimiterspace}\!\lower0.7ex
\hbox{${\scriptstyle #2}$}}}

\begin{document}

\title[Radiation and self-polarization of neutral fermions]
{Radiation and self-polarization of neutral fermions  in
quasi-classical description}

\author{A. E. Lobanov}

\address{Moscow State University, Department of Theoretical
Physics, 119992 Moscow, Russia}

\ead{lobanov@phys.msu.ru}



\begin{abstract}
A Lorentz invariant formalism for quasi-classical description of
electromagnetic radiation from a neutral spin $\oks{1}{2}$
particle with an anomalous magnetic moment moving in an external
electromagnetic field is developed. In the high symmetry fields,
for which analytical solutions to the Barg\-mann--Michel--Telegdi
equation are known, the so called self-polarization axes, i. e.
directions of preferred polarization of particles in the radiation
process, are found. Expressions for the radiative transition
probability and spectral-angular distribution of the radiation
emitted by a pola\-rized particle are obtained in the fields under
consideration.

\end{abstract}



\section{Introduction}\label{c3.b}

Electromagnetic radiation from an uncharged spin $\oks{1}{2}$
particle with an anomalous magnetic moment moving in certain types
of classical electromagnetic field was studied previous\-ly within
the Furry picture of quantum electrodynamics (QED)
\cite{1,2,3,5,6,7,60,8}. Meanwhile it was observed that under
certain conditions the radiation process may be described in
purely classical terms using the Bargmann--Michel--Telegdi (BMT)
spin evolution equation \cite{9,10}{\footnote{For more details,
see \cite{B} and references therein.}}. In such a pseudoclassical
treatment the radiation power is given by the well-known formula
for a magnetic moment radiation \cite{11}:
\begin{equation}\label{1}
  \frac {dI}{d{O}}=
 -\frac {1}{4\pi(lu)^5u^0}\{{{\ddot \mu}^\nu}
{{\ddot \mu}_\nu}(ul)^2+(l^\nu {\ddot \mu}_\nu)^2\},
\end{equation}
where $\mu^\nu$ is the 4-vector of the magnetic moment and $u^\nu$
is the 4-velocity of the particle. Here $l^\nu=\{ 1,{\bf l}\},$
where ${\bf l}$ is the unit vector in the direction of an emitted
wave; a dot denotes the differentiation with respect to the proper
time $\tau$. We use the units $\hbar =c=1 $.

The radiation power calculated within the framework of QED is
found to correspond to the result obtained from equation (\ref{1})
under the conditions of quasi-classical character of the motion,
namely, that the binding energy due to the magnetic moment in the
rest frame should be much smaller than the particle mass, and the
external field should vary slowly at the distances of the order of
the Compton length, which amounts to the conditions:
\begin{equation}\label{2}
\mu_{0}{H_0}\ll mc^{2},\quad{\hbar\dot {H}_0}/{mc^2H_0}\ll 1,
\end{equation}
\noindent where $H_0$ is the magnetic field strength in the rest
frame and $\mu_{0}$ is the value of particle  magnetic moment
(here we use Gaussian units).

In our papers \cite{L38,L36} radiation of unpolarized neutral
particles was investigated in quasi-classical approach in the case
of arbitrary field. To study radiation from an unpolarized
particle one must impose an additional requirement that consists
in averaging over the initial spin states and summing over the
final polarizations. The averaging of the quantum transition
amplitudes should correspond to the averaging over the initial
orientations of the magnetic dipole moment within the
quasi-classical consideration. We proposed \cite{L38,L36} to
replace the magnetic moment by
\begin{equation}\label{4}
\mu^\nu=\mu_0 S^\nu,
\end{equation}
where $S^\nu$ is the mean value of the spin vector, its evolution
being described by the BMT equation \cite{12}:
\begin{equation}\label{6}
\dot S^\nu=2\mu_0\{F^{\nu\alpha}S_\alpha -u^\nu (u_\alpha
F^{\alpha\beta}S_\beta)\}.
\end{equation}
\noindent The validity of this equation is ensured by the
conditions (\ref{2}) \cite{13, 14}.

Our main goal was to show that when the averaging over
polarization states at $\tau=\tau_0$ is performed, the resulting
expression for the radiation power depends only on the external
field intensity and thus is valid in the case of arbitrary
external field subjected to the conditions (\ref{2}). It is
important that the neutral particle moves with a constant velocity
in the external field. Of course, the true quantum description of
radiation demands  the accounting of quantum recoil in the photon
emission process. But when conditions (\ref{2}) are satisfied, the
energies of emitted photons are small, therefore we can neglect
the change of the particle velocity.

Unfortunately, while studying radiation of polarized particles,
even with the assumptions similar to ones discussed above, the
approach based on formula (\ref{1}) is valid only for the
transitions without spin-flip. So here we will use another method
\cite{L50} based on the introduction of quasi-classical spin wave
functions in QED formulas.

Such wave functions can be constructed as follows \cite{L32}.
Suppose the Lorentz equation is solved, i.e. the dependence of
coordinates of the particle on proper time is found. Then BMT
equation transforms to ordinary differential equation, resolvent
of which determines a one-parametric subgroup of the Lorentz
group. The quasi-classical spin wave function signifies a
spin-tensor, whose evolution is determined by the same
one-parametric subgroup.

It is easy to verify that for a neutral particle with spin
$\oks{1}{2}\,,$ represented by a Dirac bispinor, the equation for
the wave function $\Psi(\tau)$ under consideration is \cite{L32}
\begin{equation}\label{ar9}
  \dot{\Psi} = i\mu_{0}
  \gamma^{5}H^{\mu\nu}u_{\nu}\gamma_{\mu}\hat{u}\, \Psi,
\end{equation}
\noindent where $H^{\mu\nu} =
-\,\oks{1}{2}e^{\mu\nu\rho\lambda}F_{\rho\lambda}$ is the dual
electromagnetic tensor. Obviously, the density matrix of partially
polarized fermion takes the form
\begin{equation}\label{x11}
 \rho(\tau,\tau') =\frac{1}{2}\,U(\tau,\tau_{0}) \big(\hat
 {p}(\tau_{0})+m\big)\big(1-\gamma^5\hat
{S}(\tau_{0})\big)U^{-1}(\tau',\tau_{0}),
\end{equation}

\noindent where $U(\tau,\tau_{0})$ is the resolvent of equation
(\ref{ar9}). For a pure state the density matrix reduces to direct
product of bispinors, normalized by the condition
$\bar{\Psi}(\tau)\Psi(\tau) = 2m.$

\section{General relations}\label{c3.c}

Now let us investigate the radiation of polarized particles on the
base of the above considerations. The formula of quantum
electrodynamics which describes the transition probability of a
neutral fer\-mion under spontaneous radiation in external field
is{\footnote{In the expression for the radiation energy $\mathcal
E$ the additional multiple $k$ --- the energy of radiated photon
--- appears in the integrand.}}:
\begin{equation}\label{ar1}
\begin{array}{ll}
  \displaystyle P\!\!\!&=\displaystyle -\!\!\int\!\! d^{4}x\,d^{4}y\!\int
  \frac{d^{4}p\,d^{4}q\,d^{4}k}{(2\pi)^{6}}\,
\delta(k^{2})\,\delta(p^{2}-m^{2})\,
 \delta(q^{2}-m^{2})\\[10pt]&\quad\times\varrho^{\mu\nu}_{ph}(x,y;k)\:
{\mathrm{S p}}\big\{\Gamma_{\mu}(x)\varrho_{i}(x,y;p)
\Gamma_{\nu}(y)\varrho_{f}(y,x;q)\big\}.
\end{array}
\end{equation}

\noindent Here $\varrho_{i}(x,y;p),\;\varrho_{f}(y,x;q)$ are
density matrices of initial $(i)$ and final $(f)$ states of the
fermion, $\,\varrho^{\mu\nu}_{ph}(x,y;k)$ is density matrix of the
radiated photon, $\Gamma^{\mu} = -\,\sqrt{4\pi}
\mu_{0}\sigma^{\mu\nu}k_{\nu}$ is the vertex function.

In order to pass to the quasi-classical approximation, it is
necessary to substitute precise density matrices for the ones
constructed in \cite{L32}{\footnote{Note, that for arbitrary
plane-wave fields such substitution is identical.} (see
(\ref{x11})) and to neglect the recoil in the photon emission
process. The latter operation implies inserting the following
expression:
\begin{equation}\label{ar04}
(2\pi)^{3}\, \frac{p^{0}q^{0}}{m^{2}}\int\!\!\!\int\! d\tau d\tau'
\delta^{4}(x^{\alpha}-u^{\alpha}\tau)\delta^{4}
(y^{\beta}-u^{\beta}\tau')
\delta({\bf{p}}-m{\bf{u}})\delta({\bf{q}}-m{\bf{u}}),
\end{equation}
\noindent in the integrand of (\ref{ar1}), which reduces the
integration to the particle trajectory. After summation over
photon polarizations and integration with respect to fermion
momenta and coordi\-nates, we obtain the quasi-classical
expressions for the transition probability under
investi\-ga\-tion:
\begin{equation}\label{ar5}
P=\frac{\mu_{0}^{2}}{(2\pi)^{2}} \int\!\! dO\!\!
\int\limits_{0}^{\infty}\!k^{3}dk\!\int\!\!\!\int\!d\tau
d\tau'e^{ik(lu)(\tau-\tau')}\,T(\tau,\tau';u),
\end{equation}
\noindent and for the radiation energy
\begin{equation}\label{ar6}
{\mathcal{E}}=\frac{\mu_{0}^{2}}{(2\pi)^{2}} \int\!\! dO\!\!
\int\limits_{0}^{\infty}\!k^{4}dk\!\int\!\!\!\int\!d\tau
d\tau'e^{ik(lu)(\tau-\tau')}\,T(\tau,\tau';u).
\end{equation}
\noindent Here the following notation is introduced:
\begin{equation}\label{ar7}
T(\tau,\tau';u)=\left(V_{i}V_{f}-A_{i}A_{f}\right),
\end{equation}
\noindent where $V_{if},A_{if}$ are determined by formulas:
\begin{equation}\label{ar8}
\begin{array}{l}
\displaystyle  V_{i}= \frac{1}{4}{\mathrm {Sp}}\!\left\{
\hat{l}\,U(\tau)(1+\hat{u})
  (1\!-\gamma^{5}{\hat{S}}_{0i})U^{-1}(\tau')\right\}\!,\\[12pt]
\displaystyle  V_{f}= \frac{1}{4}{\mathrm {Sp}}\!\left\{
\hat{l}\,U(\tau')(1+\hat{u})
  (1\!-\gamma^{5}{\hat{S}}_{0f})U^{-1}(\tau)\right\}\!,\\[12pt]
\displaystyle A_{i}=\frac{1}{4}{\mathrm {Sp}}\Big\{\!\gamma^{5}
\hat{l}\,U(\tau)(1+\hat{u})
  (1\!-\gamma^{5}{\hat{S}}_{0i})U^{-1}(\tau')\Big\},\\[12pt]
\displaystyle A_{f}=\frac{1}{4}{\mathrm {Sp}}\Big\{\!\gamma^{5}
\hat{l}\,U(\tau')(1+\hat{u})
  (1\!-\gamma^{5}{\hat{S}}_{0f})U^{-1}(\tau)\Big\}.
\end{array}
\end{equation}
In formulas (\ref{ar5}) and (\ref{ar6})  the operator $U(\tau)
\equiv U(\tau, \tau_{0})$ is the resolvent of the equation
(\ref{ar9}). Naturally, the spectral-angular distributions for the
transition probability and radiation energy can be obtained only
if the solution of the equation (\ref{ar9}) is known. So first of
all we concentrate on investigating the general properties of
these physical values.

Let us introduce to the integration variable $k'=k(lu)$ which
denotes the photon energy in the rest frame of the radiating
particle. Using formulas for the Fourier transforms of generalized
functions (for example, see \cite{BP77B}), we integrate
(\ref{ar5}) and (\ref{ar6}) with respect to variable $k'.$ As a
result, we obtain
\begin{equation}\label{ar11}
P=\frac{\mu_{0}^{2}}{(2\pi)^{2}} \int\!
  \frac{d O}{(lu)^{4}}\!
\int\!\!\!\int\!d\tau d\tau'\frac{1}{2(\tau\!-\tau'\!+i0)}
  (\partial_{\tau}\partial_{\tau'}^{2}-
  \partial_{\tau}^{2}\partial_{\tau'})T(\tau,\tau';u),
\end{equation}
\vspace{-14pt}
\begin{equation}\label{ar12}
{\mathcal{E}} =\frac{\mu_{0}^{2}}{(2\pi)^{2}} \int\!
  \frac{d O}{(lu)^{5}}\!
\int\!\!\!\int\!d\tau d\tau'\frac{i}{(\tau\!-\tau'\!+i0)}
  \partial_{\tau}^{2}\partial_{\tau'}^{2}\,T(\tau,\tau';u).
\end{equation}

The expressions (\ref{ar11}) and (\ref{ar12}) could be integrated
with respect to the angular variables. As $U(\tau)$ is the
resolvent of the equation (\ref{ar9}), one has
\begin{equation}\label{ar012}
\begin{array}{l} \displaystyle
\int \!\frac{dO}{(lu)^{4}}\,T(\tau,\tau';u)\rightarrow
  \frac{16\pi}{3}\tilde{V}_{i}\tilde{V}_{f}, \\[12pt]
\displaystyle \int\!
\frac{dO}{(lu)^{5}}\,T(\tau,\tau';u)\rightarrow
  \frac{16\pi}{3}u^{0}\tilde{V}_{i}\tilde{V}_{f},
\end{array}
\end{equation}
\noindent where
\begin{equation}\label{ar013}
\begin{array}{l} \displaystyle
\tilde{V}_{i}= \frac{1}{4}{\mathrm {Sp}}\left\{ U(\tau)
  (1-\gamma^{5}{\hat{S}}_{0i})U^{-1}(\tau')\right\},\\[12pt]
\displaystyle \tilde{V}_{f}= \frac{1}{4}{\mathrm {Sp}}\left\{
U(\tau')
  (1-\gamma^{5}{\hat{S}}_{0f})U^{-1}(\tau)\right\}.
\end{array}
\end{equation}

Thus
\begin{equation}\label{ar0011}
  P=\frac{4\mu_{0}^{2}}{3\pi}\!
\int\!\!\!\int\!d\tau d\tau'\frac{1}{2(\tau\!-\tau'\!+i0)}
  (\partial_{\tau}\partial_{\tau'}^{2}-
  \partial_{\tau}^{2}\partial_{\tau'})\tilde{V}_{i}\tilde{V}_{f},
\end{equation}
\begin{equation}\label{ar0012}
 {\mathcal{E}} =\frac{4\mu_{0}^{2}u^{0}}{3\pi}\!
 \int\!\!\!\int\!d\tau d\tau'\frac{i}{(\tau\!-\tau'\!+i0)}\,
  \partial_{\tau}^{2}\partial_{\tau'}^{2}\tilde{V}_{i}\tilde{V}_{f}.
  \qquad
\end{equation}

Obviously, function $T(\tau,\tau';u)$  does not vary if
$\tau\leftrightarrow\tau'$ and $ S^{\mu}_{0i}\leftrightarrow
S^{\mu}_{0f}.$ It gives the possibility to obtain the following
expressions for the angular distribution
\begin{equation}\label{ar13}
\frac{d I}{d O}\Big|_ {S_{0i}\rightarrow S_{0f}}\!+\frac{d I}{d
O}\Big|_ {S_{0f}\rightarrow S_{0i}} =\frac{\mu_{0}^{2}}{2\pi u^{0}
  (lu)^{5}}\,
\partial_{\tau}^{2}\partial_{\tau'}^{2}\,
T(\tau,\tau';u)\Big|_{\tau=\tau'}\,,
\end{equation}
\noindent and for the total radiation power
\begin{equation}\label{ar14}
I_{S_{0i}\rightarrow S_{0f}}+ I_{S_{0f}\rightarrow S_{0i}}
=\frac{8\mu_{0}^{2}}{3}\,
\partial_{\tau}^{2}\partial_{\tau'}^{2}\,
\tilde{V}_{i}\tilde{V}_{f}\Big|_{\tau=\tau'}\,.
\end{equation}

\noindent If we denote the solutions of BMT equation with initial
conditions $S^{\mu}_{i }(\tau_{0})=S^{\mu}_{0 i}$ and $S^{\mu}_{f
}(\tau_{0})=S^{\mu}_{0 f}$ as $S^{\mu}_{i}$ and $ S^{\mu}_{f}$ we
obtain
\begin{equation}\label{ar15}
\begin{array}{l}
\fl\displaystyle \frac{d I}{d O}\Big|_ {S_{0i}\rightarrow
S_{0f}}\!+\frac{d I}{d O}\Big|_ {S_{0f}\rightarrow
S_{0i}}\\[4pt]\displaystyle =\frac{\mu_{0}^{2}}{2\pi u^{0}
  (lu)^{5}}
\Big\{ 2(lu)^{2}\big[4H^{2}\big((H^{2})\\[4pt]+(HS_{i})(HS_{f})\big)
-\big({\dot{H}}^{2} +({\dot{H}}S_{i})({\dot{H}}S_{f})\big)\big] \\[4pt]
-4(lu)\big[(HS_{i})e_{\mu\nu\rho\lambda}l^{\mu}{\dot{H}}^{\nu}H^{\rho}
S_{f}^{\lambda}+(HS_{f})e_{\mu\nu\rho\lambda}l^{\mu}{\dot{H}}^{\nu}H^{\rho}
S_{i}^{\lambda}\big]\\[4pt]
 -2\big[\big(4H^{2}(H l)^{2}-({\dot{H}}l)^{2}\big)+4H^{2}(H l)
e_{\mu\nu\rho\lambda}l^{\mu}{\dot{H}}^{\nu}u^{\rho}
H^{\lambda}\big]\\[2pt]-2\big[(lS_{i})(lS_{f})\big(4(H^{2})^{2}
+{\dot{H}}^{2}\big)+8(H l)^{2}(HS_{i})(HS_{f})\big]\\[4pt]
+8H^{2}(H l)\big[(lS_{i})(HS_{f})+(lS_{f})(HS_{i}) \big]\\[4pt]
+2({\dot{H}}l)\big[(lS_{i})({\dot{H}}S_{f})+(lS_{f})({\dot{H}}S_{i})\big]
\\[4pt]
-4H^{2}\big[(lS_{i})
e_{\mu\nu\rho\lambda}l^{\mu}{\dot{H}}^{\nu}u^{\rho}
S_{f}^{\lambda}+(lS_{f})
e_{\mu\nu\rho\lambda}l^{\mu}{\dot{H}}^{\nu}u^{\rho}
S_{i}^{\lambda}\big]\\[4pt]
-8H^{2}e_{\mu\nu\rho\lambda}l^{\mu}{{H}}^{\nu}u^{\rho}
S_{i}^{\lambda}\,e_{\mu\nu\rho\lambda}l^{\mu}{{H}}^{\nu}u^{\rho}
S_{f}^{\lambda}\\[4pt]
-2e_{\mu\nu\rho\lambda}l^{\mu}{\dot{H}}^{\nu}u^{\rho}S_{i}^{\lambda}\,
e_{\mu\nu\rho\lambda}l^{\mu}{\dot{H}}^{\nu}u^{\rho}S_{f}^{\lambda}
\\ -4\big((lS_{i})(H{\dot{H}})-(H l)({\dot{H}}S_{i})-
({\dot{H}}l)(HS_{i})\big)e_{\mu\nu\rho\lambda}l^{\mu}{{H}}^{\nu}u^{\rho}
S_{f}^{\lambda} \\[4pt]-4\big((lS_{f})(H{\dot{H}})-(H
l)({\dot{H}}S_{f})-
({\dot{H}}l)(HS_{f})\big)e_{\mu\nu\rho\lambda}l^{\mu}{{H}}^{\nu}u^{\rho}
S_{i}^{\lambda}\Big\},\quad\;\;
\end{array}
\end{equation}
\begin{equation}\label{ar16}
\begin{array}{l}
 \fl I_{S_{0i}\rightarrow S_{0f}}
+I_{S_{0f}\rightarrow S_{0i}} \\[4pt] =
\displaystyle\frac{16}{3}\,\mu_{0}^{2}
\Big\{4H^{2}\big((H^{2})+(HS_{i})(HS_{f})\big)-\big({\dot{H}}^{2}
+({\dot{H}}S_{i})({\dot{H}}S_{f})\big)
\\
 -2\big[(HS_{i})e_{\mu\nu\rho\lambda}u^{\mu}
 {\dot{H}}^{\nu}H^{\rho}S_{f}^{\lambda}+(HS_{f})e_{\mu\nu\rho\lambda}
 u^{\mu}{\dot{H}}^{\nu}H^{\rho}S_{i}^{\lambda}\big]\Big\},
\end{array}
\end{equation}
\noindent where $H^{\mu}$ denotes $\mu_{0}H^{\mu\nu}u_{\nu}.$ If
we average over the initial spin states and summarize over the
final ones in expressions (\ref{ar15}) and (\ref{ar16}), for which
purpose we must set $S^{\mu}_{i}=S^{\mu}_{f}=0,$  we obtain the
angular distribution and total radiation power of unpolarized
fermion \cite{L38,L36}. If we set
$S^{\mu}_{i}=S^{\mu}_{f}=S^{\mu}$ in formulas (\ref{ar15}) and
(\ref{ar16}) and divide the obtained expression by $2,$ we obtain
the angular distribution and total radiation power without spin
flip. One can see that the radiation powers without spin flip are
equal for the states of the particle with opposite polarizations.

Using the above technique, we obtain for the transition
probability at the time moment $t$:
\begin{equation}\label{ar21}
\begin{array}{l}
\fl \displaystyle \frac{d W}{d O}\Big|_ {S_{0i}\rightarrow
S_{0f}}\!\!-\frac{d W}{d O}\Big|_ {S_{0f}\rightarrow S_{0i}} \\[8pt]
=\displaystyle \frac{\mu_{0}^{2}}{4\pi u^{0}
  (lu)^{4}}
\Big\{ 4\big((HS_{i})-(HS_{f})\big)\big(H^{2}(lu)^{2}-(H
l)^{2}\big)
 \\[4pt]
-(lu)e_{\mu\nu\rho\lambda}l^{\mu}{\dot{H}}^{\nu}H^{\rho}
(S_{i}^{\lambda}-S_{f}^{\lambda})+\big((lS_{i})-(lS_{f})\big)
e_{\mu\nu\rho\lambda}l^{\mu}{\dot{H}}^{\nu}{H}^{\rho}u^{\lambda}
\\[4pt] -(H l)
e_{\mu\nu\rho\lambda}l^{\mu}{\dot{H}}^{\nu}u^{\rho}
(S_{i}^{\lambda}-S_{f}^{\lambda})+(\dot{H}l)
e_{\mu\nu\rho\lambda}l^{\mu}{H}^{\nu}u^{\rho}
(S_{i}^{\lambda}-S_{f}^{\lambda})
 \Big\},\quad\;\;
\end{array}
\end{equation}
\begin{equation}\label{ar22}
\begin{array}{l}
\fl W_{S_{0i}\rightarrow
S_{0f}}-W_{S_{0f}\rightarrow S_{0i}} \\[4pt]
=\displaystyle\frac{4\mu_{0}^{2}}{3u^{0}
  }
\Big\{ 4H^{2}\big((HS_{i})-(HS_{f})\big)
-e_{\mu\nu\rho\lambda}u^{\mu}{\dot{H}}^{\nu}H^{\rho}
(S_{i}^{\lambda}-S_{f}^{\lambda})
 \Big\}.\qquad\;\;
\end{array}
\end{equation}

Let us define the state of total self-polarization $S^{\mu}_{0tp}$
by either of the two equivalent conditions
\begin{equation}\label{ar19}
  I_{S_{0tp}\rightarrow \, -S_{0tp}}=0,\qquad W_{S_{0tp}\rightarrow \,
  -S_{0tp}}=0.
\end{equation}
\noindent In the general case, in this state
$I_{S_{0tp}\rightarrow S_{0tp}}\neq 0.$ If the state of total
self-polarization  $S^{\mu}_{0tp}$  is known, then radiation power
of the particle in $- S^{\mu}_{0tp}$ state is
\begin{equation}\label{ar20}
\begin{array}{ll}
I_{-S_{0tp}}\!\!\!&= \displaystyle\frac{8}{3}\,\mu_{0}^{2}
\Big\{4H^{2}\big(3H^{2}-(HS_{tp})^{2}\big) -\big(3{\dot{H}}^{2}-
 (\dot{H}S_{tp})^{2}\big)\\[4pt] &
 +\,4(HS_{tp})e_{\mu\nu\rho\lambda}u^{\mu}
 {\dot{H}}^{\nu}H^{\rho}S_{tp}^{\lambda}\Big\}.
\end{array}
\end{equation}

Underline that the obtained expressions do not allow to decide,
which state of the particle is the state with maximum possible
radiative self-polarization, also what is the self-polarization
degree, and whether the state of total self-polarization exists.
This is due to the fact that in  the general case the quantity  $
W_{S_{0}\rightarrow -S_{0}}+W_{-S_{0}\rightarrow S_{0}}$ necessary
for obtaining the balance equation (see, for example, \cite{ST})
could not be expressed in terms of solutions to the BMT equation.

However, we can show that there always exists either the total
self-polarization state, or the state with maximum possible
self-polarization for a neutral fermion moving in an external
field, provided that this field belongs to a class of fields
admitted, on one hand, by the requirement that equation
(\ref{ar9}) and consequently  BMT equation should allow precise
analytical solutions, and, on the other hand, specified by
conditions of solvability that do not depend on the magnitude of
the anomalous moment $\mu_{0}$. The latter can become the total
self-polarization state under continuous variation of the field
characteristics.

\section
{Radiation and self-polarization  in the fields of special type}
\label{c3.d}

Let  us prove the above statement. The conditions for the
existence of analytical solutions of BMT equation and their form
are given in \cite{L44}\footnote{The conditions for the existence
of analytical solutions of BMT equation are reduced to the
condition (\ref{ar026}) (see also \cite{L31}).}. Using the results
of \cite{L44}, in order to obtain the solutions of (\ref{ar9}), we
introduce the following basis vectors:
\begin{equation}\label{ar022}
  \begin{array}{l}
\displaystyle n^{\mu}_{0}=u^{\mu},\quad
n^{\mu}_{1}=H^{\mu}/\sqrt{-H^{2}},
\\[3pt]
\displaystyle
n^{\mu}_{2}={\big(H^{\mu}{(H\dot{H})}-\dot{H}^{\mu}H^{2}\big)}/
 {\sqrt{-H^{2}N}}, \\[3pt]
\displaystyle
n^{\mu}_{3}=-\,e^{\mu\nu\rho\lambda}u_{\nu}H_{\rho}{\dot{H}}_{\lambda}
/ \sqrt{N},
\end{array}
\end{equation}
\noindent where
\begin{equation}\label{ar23}
N={H^{2}{\dot{H}}^{2}-(H\dot{H})^{2}}.
\end{equation}
\noindent This basis is orthogonal and its elements satisfy the
system of equations, which is a four-dimensional generalization of
the Frenet equations:
\begin{equation}\label{ar24}
 {\dot{n}}^{\mu}_{1}=\kappa n^{\mu}_{2},\qquad
{\dot{n}}^{\mu}_{2}=\varkappa n^{\mu}_{3}-\kappa
n^{\mu}_{1},\qquad
  {\dot{n}}^{\mu}_{3}=-\varkappa n^{\mu}_{2}.
\end{equation}
\noindent Here parameters $\kappa$ and $ \varkappa$  are analogs
of curvature and torsion:
\begin{equation}\label{ar024}
\kappa=\frac{N^{1/2}}{(-H^2)},\qquad
 \varkappa={\frac{\displaystyle\sqrt{-H^2}}{N} }e^{\alpha\beta\gamma\delta}
{\ddot{H}_{\alpha}}{\dot{H}_{\beta}}H_{\gamma}u_{\delta}.
\end{equation}
\noindent In the chosen basis the spin vector is of the form
\begin{equation*}
S^{\mu}=\sum^{3}_{i=1}S_{i}n_{i}^{\mu},
\end{equation*}
\noindent where $S_{i}= -(Sn_{i})$ are components of
three-dimensional spin vector $\mathbf{S}.$

The $\mathbf{S}$ components satisfy the set of equations
\begin{equation}\label{ar0024}
 \begin{array}{l}
 \displaystyle \dot{S}_{1}=\kappa S_{2}, \\[2pt]
\displaystyle \dot{S}_{2}=\big(\varkappa -2
\sqrt{-H^{2}}\;\big)S_{3}-\kappa
 S_{1},\\[2pt]
\displaystyle  \dot{S}_{3}=-\big(\varkappa -2
\sqrt{-H^{2}}\;\big)S_{2}.
 \end{array}
\end{equation}
\noindent Let us define $N_{i}=\gamma^{5} \hat{n}_{i} \hat{u}.$
Obviously $N_{i}N_{j}= -ie_{ijk}N_{k},$ $\bar{N}_{i}={N}_{i}.$
Using standard transition from spinor to bispinor representation
\cite{15}, we obtain from formulas \cite{L44}
\begin{equation}\label{ar0240}
\begin{array}{l}
U=VR_{0},\\[4pt]
\displaystyle  V=\cos \frac{\Omega_{0}}{2} + i({\bf Nt}_{0}) \sin
  \frac{\Omega_{0}}{2}\,,\\[8pt]
  \displaystyle  R_{0}=\cos \frac{\Omega}{2}
   - i({\bf Nt})
   \sin
  \frac{\Omega}{2}\,.
\end{array}
\end{equation}
\noindent Here
\begin{equation}\label{ar26}
  \begin{array}{l}
\displaystyle{\bf t}_{0}
=\big\{\eta_{\varkappa},\,0,\,\eta_{\kappa}\big\}
\big(\eta_{\varkappa}^{2}+\eta_{\kappa}^{2}\big)^{-1/2},\\[3pt]
 \displaystyle
\Omega_{0}=\big(\eta_{\varkappa}^{2}+\eta_{\kappa}^{2}\big)^{1/2}\!
\int\limits_{\tau_{0}} ^{\tau}\!\sqrt{-H^{2}}\,d\tau,\\[4pt]
{\bf t} =\big\{\eta_{\varkappa}-2,\,0,\,\eta_{\kappa}\big\}
\big((\eta_{\varkappa}-2)^{2}+\eta_{\kappa}^{2}\big)^{-1/2}, \\[3pt]
\displaystyle
\Omega=\big((\eta_{\varkappa}-2)^{2}+\eta_{\kappa}^{2}\big)^{1/2}
\!\int\limits_{\tau_{0}} ^{\tau}\!\sqrt{-H^{2}}\,d\tau,
  \end{array}
\end{equation}
\noindent where
\begin{equation}\label{ar026}
\eta_{\varkappa}= \frac{\varkappa}{\sqrt{-H^{2}}}=
{\mathrm{const}},\qquad\eta_{\kappa}=
\frac{\kappa}{\sqrt{-H^{2}}}={\mathrm{const}}.
\end{equation}

The meaning of the introduced vectors ${\bf{t}}_{0}$ and
${\bf{t}}$ is quite evident. The  vector ${\bf{t}}_{0}$ represents
the axis about which the trihedron determining the field
orientation in the rest frame of the particle rotates (i. e. the
Darboux vector). Vector ${\bf{t}}$ defines the axis about which
the spin vector precesses in the particle rest frame.

Note that, if the conditions (\ref{ar026}) are satisfied, the
vector ${\bf{t}}$ is the only constant solution of the system
(\ref{ar0024}). The corresponding solution of BMT equation is:
\begin{equation}\label{ar0026}
\begin{array}{ll}
\displaystyle
S^{\mu}\!\!\!&=\left[H^{\mu}\big(\varkappa-2\sqrt{-H^{2}}\;\big)/\sqrt{-H^{2}}\right.
-\,e^{\mu\nu\rho\lambda}u_{\nu}\left.H_{\rho}{\dot{H}}_{\lambda}
\,{\kappa}/\sqrt{N}\right]\\&\times\displaystyle
\left[\big(\varkappa-2\sqrt{-H^{2}}\;\big)^{2}+{\kappa}^{2}\right]^{-1/2}.
\end{array}
\end{equation}
\noindent Characteristic of this solution is that the spin vector
(\ref{ar0026}) retains the orientation with respect to the
external field. As we shall see later, it is precisely this vector
that fixes the direction of the self-polarization axis.

Inserting (\ref{ar024}) into (\ref{ar013}), we obtain
\begin{equation}\label{ar27}
\begin{array}{ll}
 \displaystyle \tilde{V}_{i}\!\!\!&=\displaystyle
 \left\{\cos \frac{\Omega -\Omega'}{2}+
 i({\bf{t S}}_{0i})
  \sin \frac{\Omega -\Omega'}{2}\right \}\cos \frac{\Omega_{0}
  -\Omega'_{0}}{2}\\&
\displaystyle +\left\{({\bf{tt}}_{0}) \sin \frac{\Omega
-\Omega'}{2}-i\bigg[({\bf{tt}}_{0})({\bf{t S}}_{0i}) \cos
\frac{\Omega -\Omega'}{2} \right. \\&  + \left. \displaystyle
\big(({\bf S}_{0i}{\bf t})- ({\bf{tt}}_{0})({\bf{t S}}_{0i})\big)
\cos \frac{\Omega +\Omega'}{2}\right. \\& \left.+\,\,\displaystyle
({\bf S}_{0i}[{\bf t}\times {\bf t}_{0}]) \,\sin \frac{\Omega
+\Omega'}{2}\,  \,\bigg]\right\}\displaystyle \sin
\frac{\Omega_{0}
  -\Omega'_{0}}{2}\,.
\end{array}
\end{equation}}
\noindent The expression for $\tilde{V}_{f}$ is obtained from
(\ref{ar27}) after the substitution ${\bf{S}}_{0i}\!\rightarrow\!
-\,{\bf{S}}_{0f}.$ If we set ${\bf{S}}_{0i}=\zeta_{i}{\bf{t}}$ and
${\bf{S}}_{0f}=\zeta_{f}{\bf{t}},$ we obtain
\begin{equation}\label{ar28}
\begin{array}{ll}
\displaystyle
\!\!\!\!\!\tilde{V}_{i}\tilde{V}_{f}\!\!\!&=\displaystyle
\frac{1}{4}e^{i(\Omega
-\Omega')(\zeta_{i}-\zeta_{f})/2}\\[8pt]&\times\displaystyle
\left\{\left[e^{i(\Omega_{0}
  -\Omega'_{0})(\zeta_{i}+\zeta_{f})/2}\right.\right.+\left.
  e^{-i(\Omega_{0}
  -\Omega'_{0})(\zeta_{i}+\zeta_{f})/2}\right]\!\big(1-({\bf{tt}}_{0})^{2}
  \big)\\[8pt]&
\displaystyle  +\, e^{i(\Omega_{0}
  -\Omega'_{0})(\zeta_{i}-\zeta_{f})/2}
  \left(1-({\bf{tt}}_{0})\right)^{2}+\left.e^{-i(\Omega_{0}
  -\Omega'_{0})(\zeta_{i}-\zeta_{f})/2}
  \left(1+({\bf{tt}}_{0})\right)^{2}\right\}\!.
\end{array}
\end{equation}
\noindent Inserting this expression into (\ref{ar0011}) and
(\ref{ar0012}), we obtain the transition probability with, as well
as without the spin flip, and consequently the total radiation
energy. If $ \sqrt{-H^{2}}={\mathrm{const}},$ we can obtain final
expressions for the above values by locking the integration
contour in lower or upper half-plane of the complex variable
$\tau-\tau'.$  Hence, if $\dot{\Omega} > \dot{\Omega}_{0},$ we
obtain the following expression for the transition probability
with spin flip :
\begin{equation}\label{ar29}
\begin{array}{l}
 \displaystyle   W_{\zeta_{i}\rightarrow
  -\zeta_{i}}=\frac{2\mu_{0}^{2}}{3u^{0}}\,
  \Big\{2( \dot{\Omega})^{3}\left(1-({\bf{tt}}_{0})^{2}\right)
 +( \dot{\Omega}-\dot{\Omega}_{0})^{3}
 \left(1+({\bf{tt}}_{0})\right)^{2}\\[8pt]\displaystyle
 \qquad\quad\;\;
  +\,( \dot{\Omega}+\dot{\Omega}_{0})^{3}\left(1-({\bf{tt}}_{0})\right)^{2}
  \Big\}\Theta(-\zeta_{i}).
\end{array}
\end{equation}
\noindent If $\dot{\Omega} < \dot{\Omega}_{0},$ then
\begin{equation}\label{ar30}
\begin{array}{l}
\displaystyle  W_{\zeta_{i}\rightarrow
  -\zeta_{i}}=\frac{2\mu_{0}^{2}}{3u^{0}}\,
  \Big\{\Big[
  (\dot{\Omega}_{0}-\dot{\Omega})^{3}\left(1+({\bf{tt}}_{0})\right)^{2}
  \Big]\Theta(\zeta_{i}) \\[8pt]
\displaystyle \qquad\quad\;\; + \Big[2(
\dot{\Omega})^{3}\left(1-({\bf{tt}}_{0})^{2}\right)+
  (\dot{\Omega}+\dot{\Omega}_{0})^{3}\left(1-({\bf{tt}}_{0})\right)^{2}
  \Big]\Theta(-\zeta_{i})\Big\}.
\end{array}
\end{equation}
\noindent Here $\Theta(x)$ is the Heaviside theta-function.

Thus if $\dot{\Omega}>\dot{\Omega}_{0}$ the state with spin vector
(\ref{ar0026}) will be the state with total self-polarization,
otherwise it will have partial self-polarization. Using
(\ref{ar27}), it is easy to verify that the maximum possible
degree of self-polarization is obtained in the state
(\ref{ar0026}):
\begin{equation}\label{ar31}
\begin{array}{l}
\fl\displaystyle \frac{W_{\zeta_{p}\rightarrow -\zeta_{p}}
  -W_{-\zeta_{p}\rightarrow \zeta_{p}}}{W_{\zeta_{p}\rightarrow -\zeta_{p}}
  +W_{-\zeta_{p}\rightarrow \zeta_{p}}}\\[12pt]
\displaystyle \!\!\!\!\!\!= \frac{2(
\dot{\Omega})^{3}\left(1-({\bf{tt}}_{0})^{2} \right)
 +( \dot{\Omega}-\dot{\Omega}_{0})^{3}\left(1+({\bf{tt}}_{0})\right)^{2}
  +( \dot{\Omega}+\dot{\Omega}_{0})^{3}\left(1-({\bf{tt}}_{0})\right)^{2}
}{2( \dot{\Omega})^{3}\left(1-({\bf{tt}}_{0})^{2}\right)
 +{|\dot{\Omega}-\dot{\Omega}_{0}|}^{3}\left(1+({\bf{tt}}_{0})\right)^{2}
  +(
  \dot{\Omega}+\dot{\Omega}_{0})^{3}\left(1-({\bf{tt}}_{0})
  \right)^{2}}
\end{array}
\end{equation}

The transition probability without spin flip is determined as:
\begin{equation}\label{ar32}
 W_{\zeta_{i}\rightarrow
  \zeta_{i}}=\frac{2\mu_{0}^{2}}{3u^{0}}\,
  \Big\{( \dot{\Omega}_{0})^{3}\left(1-({\bf{tt}}_{0})^{2}\right)
  \Big\}.
\end{equation}

 The total radiation power is
\begin{equation}\label{ar33}
\begin{array}{ll}
 \displaystyle   I_{\zeta_{i}\rightarrow
  -\zeta_{i}}\!\!\!& =\displaystyle\frac{2\mu_{0}^{2}}{3}\,
  \Big\{2(
  \dot{\Omega})^{4}\left(1-({\bf{tt}}_{0})^{2}\right)\\[8pt]&
 \displaystyle   +( \dot{\Omega}-\dot{\Omega}_{0})^{4}
 \left(1+({\bf{tt}}_{0})\right)^{2}
  +( \dot{\Omega}+\dot{\Omega}_{0})^{4}\left(1-({\bf{tt}}_{0})\right)^{2}
  \Big\}\Theta(-\zeta_{i})
\end{array}
\end{equation}
\noindent if $\dot{\Omega} > \dot{\Omega}_{0}$ and
\begin{equation}\label{ar34}
\begin{array}{ll}
\displaystyle  I_{\zeta_{i}\rightarrow
  -\zeta_{i}}\!\!\!& =\displaystyle \frac{2\mu_{0}^{2}}{3}\,
  \Big\{\Big[
  (\dot{\Omega}_{0}-\dot{\Omega})^{4}\left(1+({\bf{tt}}_{0})\right)^{2}
  \Big]\Theta(\zeta_{i}) \\[8pt]&
\displaystyle  +\Big[2(
\dot{\Omega})^{4}\left(1-({\bf{tt}}_{0})^{2}\right)+
  (\dot{\Omega}+\dot{\Omega}_{0})^{4}\left(1-({\bf{tt}}_{0})\right)^{2}
  \Big]\Theta(-\zeta_{i})\Big\}
\end{array}
\end{equation}
\noindent if $\dot{\Omega} < \dot{\Omega}_{0};$
\begin{equation}\label{ar35}
 I_{\zeta_{i}\rightarrow
  \zeta_{i}}\quad=\frac{2\mu_{0}^{2}}{3}\,
  \Big\{( \dot{\Omega}_{0})^{4}\left(1-({\bf{tt}}_{0})^{2}\right)
  \Big\}.
\end{equation}

These formulas show that in the fields under investigation, in the
rest frame, the particle can radiate photons of only four
energies: the ones corresponding to the characteristic frequency
of the external field variation, to the frequency of spin
precession, and to two combination frequencies, the radiation with
the external field frequency being possible only without spin
flip.

The obtained formulas may be somewhat simplified using explicit
expressions for $\dot{\Omega},\, \dot{\Omega}_{0},
\,{\bf{t}}_{0},$ and $\, {\bf{t}}.$ We perform this for special
fields.

\section{Examples}\label{c3.e}

First of all, we consider the case of constant homogeneous
magnetic field. This problem is in a sense a test for calculation
techniques. It was first discussed in \cite{1} within the
framework of quantum theory, and then was repeatedly investigated
using various quasi-classical methods \cite{BTB95} (see, also,
\cite{J,BBT93}). Since in the field under consideration $\kappa =
\varkappa =0,$ the integrals over $\tau ,\;\tau'$ in formulas
(\ref{ar5}), (\ref{ar6}) can be calculated precisely for arbitrary
$S_{0i},\;S_{0f}.$ We obtain \vspace{-1pt}
\begin{equation}\label{rar1}
\begin{array}{ll}
\displaystyle \frac{d W}{dO\,d k}\Big|_ {S_{0i}\rightarrow
S_{0f}}\!\!\!&= \displaystyle \frac{\mu_{0}^{2}\,k^{3}}{8\pi
u^{0}}\;\delta\big(k(lu)-2\sqrt{-H^{2}}\;\big) \\[12pt]&
\times\! \displaystyle \left(\!(lu)^{2}- \frac{(H
l)^{2}}{H^{2}}\right)\!\!\left(\!1-\frac{(HS_{0i})}{\sqrt{-H^{2}}}\right)\!\!
\left(\!1+\frac{(HS_{0f})}{\sqrt{-H^{2}}}\right)\!,
\end{array}
\end{equation}\vspace{-10pt}
\begin{equation}\label{rar2}
\begin{array}{ll}
\displaystyle \frac{d I}{dO\,d k}\Big|_ {S_{0i}\rightarrow
S_{0f}}\!\!\!&= \displaystyle \frac{\mu_{0}^{2}\,k^{4}}{8\pi
u^{0}}\;\delta\big(k(lu)-2\sqrt{-H^{2}}\;\big)\\[12pt]& \times\!
\displaystyle \left(\!(lu)^{2}- \frac{(H
l)^{2}}{H^{2}}\right)\!\!\left(\!1-\frac{(HS_{0i})}{\sqrt{-H^{2}}}\right)\!\!
\left(\!1+\frac{(HS_{0f})}{\sqrt{-H^{2}}}\right)\!.
\end{array}
\end{equation}

Obviously, in this case the self-polarization is total, and
self-polarization axis is $S^{\mu}_{tp}=
-{H^{\mu}}/{\sqrt{-H^{2}}},$ i. e., depending on the sign of
anomalous magnetic moment, in the rest frame the particle spin is
oriented either along or opposite to the direction of the magnetic
field. Because of the relation $\dot{\Omega}_{0}=0,$ the radiation
frequency in the rest frame is equal to the frequency of spin
precession. It is highly important that, under transition from any
spin state, the spectral-angular distribution of the radiation is
the same irrespective of whether the spin flip takes place. In
this case the classical formula for magnetic dipole radiation is
valid only for the transitions without spin flip. However, due to
the above feature, the radiation power calculated using this
formula differs from the correct value only in numerical
coefficient.

After the integration with respect to the angles and photon
energies, we obtain
\begin{equation}\label{rar3}
W_{S_{0i}\rightarrow S_{0f}}= \frac{16\mu_{0}^{2}}{3 u^{0}}\big(-
H^{2}\big)^{3/2}\!\left(\!1-\frac{(HS_{0i})}{\sqrt{-H^{2}}}\right)\!\!
\left(\!1+\frac{(HS_{0f})}{\sqrt{-H^{2}}}\right)\!,\\[4pt]
\end{equation}
\begin{equation}\label{rar4}
I_ {S_{0i}\rightarrow S_{0f}}\;\;= \frac{32\mu_{0}^{2}}{3 }\big(
H^{2}\big)^{2}\!\left(\!1-\frac{(HS_{0i})}{\sqrt{-H^{2}}}\right)\!\!
\left(\!1+\frac{(HS_{0f})}{\sqrt{-H^{2}}}\right)\!.\!\!\!\!
\end{equation}
\noindent If we introduce $S_{0i}^{\mu}=\zeta_{i}S^{\mu}_{tp},\;
S_{0f}^{\mu}=\zeta_{f}S^{\mu}_{tp}$ and consider the transitions
only between those states, we obtain the expressions analogous to
the ones derived in \cite{1}.

Let us now discuss the radiation in the field of a circularly
polarized monochromatic plane wave with the frequency $\omega$ and
the amplitude $E.$ In this case
\begin{equation}\label{rar7}
\begin{array}{ll}
H^{\mu}&
=-\mu_{0}E\left[\big((nu)a_{1}^{\mu}-(a_{1}u)n^{\mu}\big)\cos
\omega (n x)\right. \\[4pt]&\quad\, +
{g}\left.\big(a_{2}^{\mu}(nu)-(a_{2}u)n^{\mu}\big)\sin \omega (n
x)\right],\\[4pt] \!\!\!\!\!-H^{2}\!\!\!&= \mu_{0}^{2}(nu)^{2}E^{2},\qquad
\kappa=\omega(nu),\qquad \varkappa=0.
\end{array}
\end{equation}
\noindent Here
\begin{equation}\label{x18}
\begin{array}{ll}
 n^{\mu} = \left( {1,\;{\bf n}}\right),&\quad n^{\mu}_{ +}  =
\oks{{1}}{{2}}\left( {1, - \,{\bf n}}  \right),\\[4pt]
 a^{\mu}_{1} = \left( {0,\;{\bf a}_{1}}  \right),
 &\quad a^{\mu}_{2} = \left(
{0,\;{\bf a}_{2}}  \right),
\end{array}
\end{equation}
\noindent where $\omega{\bf{n}}$ is the wave vector and ${\bf {a}
}_{i}$ are the unit vectors of polarization.

Consequently,
\begin{equation}\label{rar8}
  {\dot\Omega}_{0}=\omega (nu) ,\quad {\dot\Omega}=\omega
  (nu)(1+d^{2})^{1/2},\quad ({\bf t}{\bf t}_{0} )=(1+d^{2})^{-1/2},
  \end{equation}
where $d=2\mu_{0}E/\omega .$ Since the condition
$\dot{\Omega}_{0}<\dot{\Omega}$ is satisfied for any wave
parameters, the transition probability is determined by
(\ref{ar29}) and (\ref{ar32}), and radiation power by (\ref{ar33})
and (\ref{ar35}).

 It was indicated above that in the particle rest frame the
photons with only four frequencies are radiated. Partial
transition probabilities and radiation powers are expressed as
\begin{equation}\label{rar9}
\begin{array}{ll}
 \displaystyle   W_{\zeta_{i}\rightarrow
  -\zeta_{i}}( \dot{\Omega})&=\displaystyle  \frac{4\mu_{0}^{2}
  \omega^{3}(nu)^{3}d^{2}}{3u^{0}}\,
  (1+d^{2})^{1/2}\Theta(-\zeta_{i}),
  \\[6pt]
\displaystyle   W_{\zeta_{i}\rightarrow
  -\zeta_{i}}( \dot{\Omega} \pm \dot{\Omega}_{0})\!\!\!&=\displaystyle
  \frac{2\mu_{0}^{2}\omega^{3}(nu)^{3}d^{4}}{3u^{0}(1+d^{2})}\,
  \big((1+d^{2})^{1/2} \pm 1 \big)\Theta(-\zeta_{i}),\\[10pt]
\displaystyle   W_{\zeta_{i}\rightarrow
  \zeta_{i}}( \dot{\Omega}_{0})&=\displaystyle  \frac{2\mu_{0}^{2}
  \omega^{3}(nu)^{3}d^{2}}
  {3u^{0}(1+d^{2})}\,;
\end{array}
\end{equation}

\begin{equation}\label{rar10}
\begin{array}{ll}
\displaystyle I_{\zeta_{i}\rightarrow
  -\zeta_{i}}( \dot{\Omega})&=\displaystyle \frac{4\mu_{0}^{2}
  \omega^{4}(nu)^{4}d^{2}}{3}\,
  (1+d^{2})\Theta(-\zeta_{i}),
  \\[6pt]
\displaystyle I_{\zeta_{i}\rightarrow
  -\zeta_{i}}( \dot{\Omega} \pm \dot{\Omega}_{0})\!\!\!&=\displaystyle
  \frac{2\mu_{0}^{2}\omega^{4}(nu)^{4}d^{4}}{3(1+d^{2})}\,
  \big((1+d^{2})^{1/2} \pm 1 \big)^{2}\Theta(-\zeta_{i}),\\[10pt]
\displaystyle   I_{\zeta_{i}\rightarrow
  \zeta_{i}}( \dot{\Omega}_{0})&=\displaystyle
  \frac{2\mu_{0}^{2}\omega^{4}(nu)^{4}d^{2}}
  {3(1+d^{2})}\,.
\end{array}
\end{equation}

The fact, that the particle can radiate not only with the
frequency of external wave, but also with other three frequencies,
which are not  multiples of the first one, was mentioned in
\cite{5}. We must emphasize, that in our case the
self-polarization axis does not determine a constant space
direction, but is rigidly tied to the external field. Namely, in
the rest frame of the particle its spin vector precesses with the
wave frequency around the wave vector, the spin vector being in
the same plane with the wave vector and the vector of magnetic
field strength, the angle between the spin vector and the wave
vector being  equal to arc tangent of parameter $ d.$

The interesting case is the Redmond field, which is the
superposition of the above discussed circularly polarized wave and
a constant homogeneous magnetic field directed along its wave
vector. Since in the Redmond field the conditions (\ref{ar026})
are satisfied only when the particle moves along the constant
magnetic field $H_{\parallel},$  we study this case and set
$(a_{i}u)=0.$ Then
\begin{equation}\label{rar11}
\begin{array}{ll}
H^{\mu} \!\!\!&=-\mu_{0}E(nu)\big[a_{1}^{\mu}\cos \omega (n x)+
{g}a_{2}^{\mu}\sin \omega (n x)\big]\\[6pt]&
-\mu_{0}H_{\parallel}\left[n^{\mu}(n_{+}u)-n^{\mu}_{+}(nu)\right],\\[6pt]
\!\!\!\!\!-H^{2} & = \mu_{0}^{2}\big((nu)^{2}E^{2}+H_{\parallel}^{2}\big), \\[6pt]
\end{array}
\end{equation}
\begin{equation}\label{rarq11}
\displaystyle \kappa =\displaystyle \frac{\omega(nu)}{(1+
H_{\parallel}^{2}/E^{2}(nu)^{2})^{1/2}}\,,\qquad \varkappa = -
\frac{{g}\mu_{0}\omega H_{\parallel}/|\mu_{0}|E}{(1+
H_{\parallel}^{2}/E^{2}(nu)^{2})^{1/2}}\,.
\end{equation}

\noindent Therefore,\vspace{-4pt}
\begin{equation}\label{rar12}
\begin{array}{ll}
\displaystyle  {\dot\Omega}_{0}&=\omega (nu) ,\\[6pt]
{\dot\Omega}&=\omega
  (nu)\big((1+2{g}\mu_{0}H_{\parallel}/\omega(nu))^{2}+d^{2}\big)
  ^{1/2},\\[6pt]
\displaystyle \!\!\!\!({\bf{tt}}_{0})\!\!&=\displaystyle
\frac{1+2{g}\mu_{0}H_{\parallel}/\omega(nu)} {\big((1+2g
\mu_{0}H_{\parallel}/ \omega(nu))^{2} + d^{2}\big)^{1/2}}\,.
\end{array}
\end{equation}

If $\varkappa < \sqrt{-H^{2}},$ then the condition
$\dot{\Omega}_{0}<\dot{\Omega}$ is satisfied. In this case, the
transition probability is defined by (\ref{ar29}) and
(\ref{ar32}), and the radiation power is defined by (\ref{ar33})
and (\ref{ar35}).

If $\varkappa > \sqrt{-H^{2}},$ then the condition
$\dot{\Omega}_{0}> \dot{\Omega}$ is satisfied. In this case the
transition probability is defined by (\ref{ar30}) and
(\ref{ar32}), whereas the radiation power is determined by
(\ref{ar34}) and (\ref{ar35}). For this situation to take place,
it is necessary that $d\leqslant1$ and, consequently, the constant
magnetic field strength should satisfy the condition:
\begin{equation}\label{rar13}
\displaystyle q_{1} < -2{g}\mu_{0}H_{\parallel}/\omega(nu) <
q_{2},
\end{equation}
\noindent where $q_{1,2}=1 \mp \sqrt{1-d^{2}}.$

The  resonant  case, where $\varkappa = \sqrt{-H^{2}},$ i. e. the
condition
\begin{equation}\label{rar15}
-2{g}\mu_{0}H_{\parallel}/\omega(nu) = q_{1,2}
\end{equation}
\noindent is satisfied, is very interesting. In this case the
radiation is possible only with the frequency of external wave and
with the double frequency.

\begin{figure}[h]
\begin{center}
\includegraphics[width=0.7\textwidth]{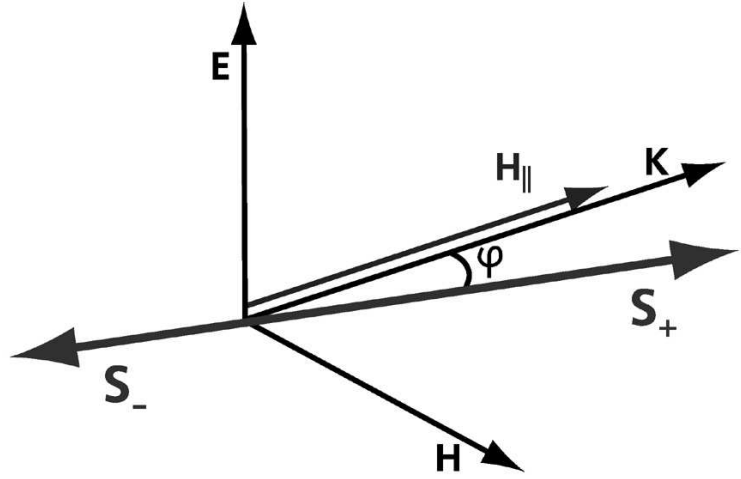}
\caption{Orientation of the self-polarization axis for the Redmond
field.}\label{bbb}
\end{center}
\end{figure}

The above formulas are illustrated by Figure \ref{bbb}. In the
Redmond field one has ${\bf t}_{0}=g{\bf n},$ hence the
self-polarization axis precesses about wave vector of
electromagnetic wave by the angle
\begin{equation}\label{rarq15}
\tan \varphi = \frac{g|d|}{1+2g\mu_{0}H_{\parallel}/\omega(nu)}.
\end{equation}
If the condition $\sin \varphi < |d|$ is valid, self-polarization
is total, hence the particle only scatters the external wave. If
$\sin \varphi
> |d|$ radiative transitions between
states with positive and negative projections on this axis exist,
frequencies of transitions being different.

It must be emphasized that the formulas for the probability of the
radiative transition and radiation power are based on the
solutions of equation (\ref{ar9}). Obviously, the above deductions
will be also true if we replace tensor $H^{\mu\nu}$ by any
antisymmetric tensor. In \cite{L43} the spin evolution equation
was deduced for a neutrino, --- a particle involved in weak
interaction. This equation possesses the same structure as BMT
equation, it can be obtained by the substitution of the
electromagnetic field tensor $F_{\mu\nu}$ in the following way:
\begin{equation}\label{w2}
F_{\mu\nu} \rightarrow E_{\mu\nu}= F_{\mu\nu}+G_{\mu\nu}.
\end{equation}
The tensor $G_{\mu\nu}$ describes the coherent interaction of
neutrino with moving and polarized matter. In the general case of
neutrino interacting with  background fermions $f$ we have
\begin{equation}\label{w3}
G^{\mu \nu}= \epsilon ^{\mu \nu \rho \lambda}
g^{(1)}_{\rho}u_{\lambda}- (g^{(2)\mu}u^\nu-u^\mu g^{(2)\nu}).
\end{equation}
Here
\begin{equation}\label{w4}
g^{(1)\mu}=\sum_{f}^{} \rho ^{(1)}_f j_{f}^\mu +\rho ^{(2)}_f
\lambda _{f}^{\mu}, \ \ g^{(2)\mu}=\sum_{f}^{} \xi ^{(1)}_f
j_{f}^\mu +\xi ^{(2)}_f \lambda _{f}^{\mu},
\end{equation}
where $j_{f}^\mu$ are fermion currents and $\lambda^{\mu}_f$ are
fermion polarizations (summation is performed over all fermions
$f$ of the background). The explicit expressions for the
coefficients $\rho_{f}^{(1),(2)}$ and $\xi_{f}^{(1),(2)}$ could be
found if a special model of neutrino interaction is chosen.

So if, in equation (\ref{ar9}), we replace $H^{\mu\nu}$ by the
tensor $Z^{\mu\nu}=
-\oks{1}{2}e^{\mu\nu\rho\lambda}E_{\rho\lambda}$, solutions of
such equation may be used for determining the intensity of
neutrino spin light in matter  and the probability of neutrino
radiative transition. Just this method was used in \cite{L49,L51}
for calculating characteristics of spin light within the
quasi-classical approximation. If the matter density is assumed to
be constant, which implies that $Z^{\mu\nu}$ is coordinate
independent, it is possible to obtain the formulas for processes
in matter and external constant electromagnetic fields, for
example, in magnetized plasma. The results can be found by the
substitution
\begin{equation}\label{w5}
H^{\mu}\rightarrow
Z^{\mu}=H^{\mu}+\mu_{0}\big({g}^{(1)\mu}
-u^{\mu}({g}^{(1)}u)\big)
\end{equation}
in equations (\ref{rar1})
-- (\ref{rar4}), obtained for the radiation in a magnetic field.

It should be emphasized that the results obtained in this section
agree in the quasi-classical region with those obtained by the
methods of quantum electrodynamics in the cases when such
calculations were carried out \cite{1,2,3,5,6,7,L58}.

It is of interest to compare the orders of magnitude of the
radiation power $I$ of a neutral particle and the classical
radiation power of a charged particle  $I_{0}$ (see, for instance,
\cite{11}).  If these particles have close values of mass, e.g., a
proton and a neutron, it is easy to find that
 \begin{equation}\label{w6}
   \frac{I}{I_0} \sim \left\{ \left( \frac {H_0}{H_{cr}}\right)^2,
\left(\frac { \hbar \dot{H}_{0}}{m c^2 H_{0}}\right)^2 \right\},
\end{equation}
where the notations are the same as in eq. (\ref{2}).

Therefore, the radiation powers of a charged particle and that of
a neutral particle have the same orders of magnitude either in
superstrong fields, which can exist in the vicinity of
astrophysical objects of pulsar type, or in very high-frequency
fields.

\section{Conclusions}\label{c3.g}

Two important conclusions follow from the obtained results. First
of all, neutral particles can emit radiation without spin flip.
This possibility depends on the following circumstance. Particle
polarization is well defined in the rest frame. To determine
polarization in the laboratory frame it is necessary to make
Lorentz transformation along the kinetic momentum of the particle.
However, for the particle in an external field, directions of
kinetic and canonical momenta, generally speaking, are different.
That is why radiative transitions without spin flip do not
contradict conservation laws. Therefore commonly used description
of radiative transitions of neutral particles as transitions with
spin flip is perfectly true only when transitions from the state
which is opposite to the total self-polarization state are
considered.

As a result of radiative transition, the particle always goes to
the state with the spin parallel to the self-polarization axis. So
even for the process in homogeneous magnetic field, the transition
probability divides into transition probabilities with and without
spin flip. When the particle moves in inhomogeneous, or especially
in non-stationary electromagnetic field, the radiative transition
without spin flip occurs even from the total self-polarization
state.  For example, this effect is inherent in the process in the
plane-wave field. In this case the particle taken in the state of
total polarization emits radiation with the frequency of the
external wave (i. e. it scatters the external wave).

In the second place, there are configurations of external fields
for which total polarization states of the particles do not exist.
The example of such configuration is the Redmond field.

\ack

The author is grateful to V.G.~Bagrov, A.V.~Borisov,
O.S.~Pav\-lova, A.E.~Shabad, and V.Ch.~Zhu\-kovsky for fruitful
discussions.

\bigskip

\noindent This work was supported in part by the grant of
President of Russian Federation for leading scientific schools
(Grant SS --- 2027.2003.2).

\end{document}